\documentclass[aps,prd,amssymb,groupedaddress,nofootinbib]{revtex4}

\usepackage{epsfig}
\usepackage[english]{babel}

\usepackage{amsfonts}
\usepackage{amssymb}
\usepackage{amsmath}
\usepackage{slashbox}
\usepackage{multirow}

\def\a{\alpha}
\def\b{\beta}

\def\d{\delta}

\def\l{\lambda}

\def\O{\Phi}

\def\p{\partial}

\def\t{\tau}

\def\w{\omega}

\def\={\nonumber &=}

\def\&{{}&}

\def\({\left(}
\def\){\right)}
\def\[{\left[}
\def\]{\right]}
\def\<{\left\langle}
\def\>{\right\rangle}

\def\uk{{\bf \hat{k}}}

\def\ux{{\bf \hat{x}}}
\def\bk{{\bf k}}

\def\bx{{\bf x}}
\def\bK{{\bf K}}
\def\by{{\bf y}}

\def\curl{\mathcal}

\def\eq{\begin{align}}
\def\qe{\end{align}}
\def\eqa{\begin{eqnarray}}
\def\qea{\end{eqnarray}}

\def\and{\quad \mbox{and} \quad}

\def\fnl{f_\textrm{NL}}

\def\Fnl{ F_\textrm {NL}}

\def\bfnl{\kern2pt\overline{\kern-2ptf}_\textrm{NL}}

\def\lmax{l_\textrm{max}}

\def\Blll{B_{l_1l_2l_3}}

\def\kall{k_1,k_2,k_3}

\def\Bkkk{B(\kall)}

\def\Qn{\curl{Q}_n}
\def\Qm{\curl{Q}_m}

\def\Rn{\curl{R}_n}
\def\Rm{\curl{R}_m}

\def\barQ{\kern2pt\overline{\kern-2pt\curl{Q}}}

\def\barR{\kern2pt\overline{\kern-2pt\curl{R}}}

\def\nmax{n_\textrm{max}}

\def\aR{\alpha^{\scriptscriptstyle{\cal R}}}
\def\aRn{\aR_n}

\def\aQ{\alpha^{\scriptscriptstyle{\cal Q}}}
\def\aQn{\aQ_n}

\def\baQ{\bar{\alpha}^{\scriptscriptstyle{\cal Q}}}

\def\bR{\beta^{\scriptscriptstyle{\cal R}}}
\def\bRn{\bR_n}

\def\bQ{\beta^{\scriptscriptstyle{\cal Q}}}
\def\bQn{\bQ_n}

\def\bbQ{\bar{\beta}^{\scriptscriptstyle{\cal Q}}}

\def\setsize{\csname @setfontsize\endcsname \setsize}

\begin{document}

%%%%%%%%%%%%%%%%%%%%%%%%%%%%%%%%% INTRODUCTION %%%%%%%%%%%%%%%%%%%%%%%%%%%%%%%%

\title{Rapid Separable Analysis of Higher Order Correlators in Large Scale Structure}

\author{J.R.~Fergusson}

\author{D.M.~Regan}

\author{E.P.S.~Shellard}

\affiliation
{Centre for Theoretical Cosmology,\\
Department of Applied Mathematics and Theoretical Physics,\\
University of Cambridge,
Wilberforce Road, Cambridge CB3 0WA, United Kingdom}

\date{\today}

\begin{abstract}
We present an efficient separable approach to the estimation and reconstruction of the bispectrum and the trispectrum from observational (or simulated) large scale structure data.  This is developed from general CMB (poly-)spectra methods which exploit the fact that the bispectrum and trispectrum in the literature can be represented by a separable mode expansion which converges rapidly (with $\nmax={\cal{O}}(30)$ terms).  With an effective grid resolution $\l_{max}$ (number of particles/grid points $N=\lmax^3$), we present a bispectrum estimator which requires only ${\cal O}(\nmax \times \lmax^3)$ operations, along with a corresponding method for direct bispectrum reconstruction. This method is extended to the trispectrum revealing an estimator which requires only  ${\cal O}(\nmax^{4/3} \times \lmax^3)$ operations. The complexity in calculating the trispectrum in this method is now involved in the original decomposition and orthogonalisation process which need only be performed once for each model. However, for non-diagonal trispectra these processes present little extra difficulty and may be performed in ${\cal O}(\lmax^4)$ operations. A discussion of how the methodology may be applied to the quadspectrum is also given. An efficient algorithm for the generation of arbitrary nonGaussian initial conditions for use in N-body codes using this separable approach is described. This prescription allows for the production of nonGaussian initial conditions for arbitrary bispectra and trispectra. A brief outline of the key issues involved in parameter estimation, particularly in the non-linear regime, is also given. 

\end{abstract}

%\pacs{1}

\maketitle

%%%%%%%%%%%%%%%%%%%%%%%%%%%%%%%%% INTRODUCTION %%%%%%%%%%%%%%%%%%%%%%%%%%%%%%%%
%\twocolumngrid
\setsize{10}{12}

\section{Introduction}

In previous work \cite{FLS09, FLS10, RSF10} we developed and implemented a methodology for the efficient and  
general analysis of nonGaussianity in the cosmic microwave sky.   Our purpose here is to apply  
these separable mode methods to  large-scale structure, making tractable a fast general 
analysis of all bispectra and trispectra, rather than the few special cases studied to date.   
Calculation of the three-point correlator or bispectrum $\langle \delta_{{\bf k}_1}\delta_{{\bf k}_2}\delta_{{\bf k}_3}\rangle$
using 3D large-scale structure data naively appears to require a computationally
intensive $\lmax^6$ operations, or $\lmax^9$ for the trispectrum, where $\lmax$ is the 
effective observational or simulated grid resolution (i.e.\ the volume sidelength $L$ over the averaged 
galaxy or grid spacing $\Delta x$, giving a particle number $N\approx\lmax^3$). However, if - as in the
CMB - predicted nonGaussianity can be described by rapidly convergent and separable mode expansions, then
there is a dramatic reduction  to only ${\cal O}(\nmax\times\lmax^3)$ operations for estimating any bispectrum, 
where $\nmax$ is the (small) number of modes required for an accurate representation ($\nmax \approx 30$ for
WMAP analysis \cite{FLS10}).   The relative impact on trispectrum estimation is even more dramatic, reducing 
again to $\sim {\cal O}(\nmax^{4/3} \times \lmax^3)$ operations.   
Direct reconstruction of the bispectrum today then allows for the decomposition into its 
constituent  and independent shapes, including contributions directly from the primordial bispectrum, 
from next-to-leading order terms in nonlinear gravitational collapse, from the convolved primordial 
trispectrum, etc.   These methods equally can be applied to 
generating simulation initial conditions with arbitrary given bispectrum and trispectrum, 
again using a simple separable mode algorithm requiring only ${\cal O}(\nmax\times\lmax^3)$ or ${\cal O}(\nmax^{4/3}\times\lmax^3)$ operations respectively.

Our purpose here is not to review the many important contributions made to the study of higher-order correlators 
in large-scale 
structure, for which there are some comprehensive recent reviews available (\cite{LigSef2010,DS}).   However, we
note that the 
field is well-motivated because nonGaussianity is recognised as a critical test of the simplest standard inflationary 
scenario.   Moreover, there are a growing number of alternative inflationary scenarios where deviations from nonGaussianity can be large (see \cite{Chen2010} for a review). The most stringent constraints on primordial nonGaussianity so far have come from CMB bispectrum measurements (e.g.\ \cite{WMAP7, FLS10}, see \cite{LigSef2010}) with relatively weak constraints coming from the large-scale structure galaxy bispectrum \cite{0312286} due to complications in dealing with non-linear evolution. While it appears to be possible also to derive competitive constraints using the abundance 
of rare objects or scale-dependent bias (e.g.\ \cite{08053580}), these complementary approaches generally assume 
a local-type nonGaussianity (see the review \cite{Verde2010}).   With improving galaxy and other surveys 
covering a growing fraction of the sky, it is reasonable to expect measurements of higher order correlators from this three-dimensional data to provide the best and most comprehensive information about nonGaussianity.   
These large-scale structure (poly-)spectra should allow us to discriminate between different non-Gaussian shapes, 
notably between primordial and late-time sources, ultimately complementing 
CMB measurements and exceeding them in precision.

In this paper we present a method for quickly calculating the bispectrum from a given density perturbation in section \ref{sec:bispectrum}.  Next we show how to extend this analysis to the trispectrum in section \ref{sec:trispectrum}. As any estimator would require nonGaussian simulations for testing and error analysis we present an approach in section \ref{sec:intcon} for including a general bispectrum and trispectrum in the initial conditions for $N$-body simulations. We then go on to show in section \ref{sec:fnlestimation} how a general estimator for constraining primordial nonGaussianity can be constructed, when the bispectrum can be approximated using a simple ansatz, and in the completely general case. Finally we present our concluding remarks.

\section{LARGE-SCALE STRUCTURE BISPECTRUM CALCULATION}\label{sec:bispectrum}

\subsection{General bispectrum estimator}

Higher-order correlators of the galaxy or matter density distribution can be expected to exhibit a low signal-to-noise
for individual combinations of wavenumbers (as for multipoles in the CMB).  A useful strategy for the comparison 
between observations and theoretical models (or simulated numerical models) is the use of an estimator which 
tests for consistency by summing over all multipoles using an optimal signal-to-noise weighting.  
The general estimator for the galaxy or density bispectrum, when searching for a given theoretical 
three-point correlator $\<\d_{\bk_1}\d_{\bk_2}\d_{\bk_3}\>$, is
\begin{align}\label{eq:bestimatorgen}
\curl{E} = \int \frac{d^3k_1}{(2\pi)^3} \frac{d^3k_2}{(2\pi)^3} \frac{d^3k_3}{(2\pi)^3} 
\<\d_{\bk_1}\d_{\bk_2}\d_{\bk_3}\>
\left[C^{-1}(\d^{obs}_{\bk_1})C^{-1}(\d^{obs}_{\bk_2})C^{-1}(\d^{obs}_{\bk_3}) - 
3 C^{-1}(\d^{obs}_{\bk_1}\d^{obs}_{\bk_2})C^{-1}(\d^{obs}_{\bk_3})\right]
\end{align}
where $\delta^{obs}_{{\bf k}}$ represents a noisy measurement of the galaxy or density perturbation with 
signal plus noise covariance $C$ given by 
\begin{align}
C^{-1}(\d^{obs}_{\bk}) = \int \frac{d^3k'}{(2\pi)^3} \<\d_{\bk}\d_{\bk'}\>^{-1}\d^{obs}_{\bk'}\,,
\end{align}
we will discuss the normalisation necessary for parameter estimation in section~\ref{sec:fnlestimation}. Here, we have added a linear term to the cubic estimator in order to account for inhomogeneous effects from 
incomplete survey coverage (e.g.\ due to dust extinction), sampling bias, shot noise, 
and other known systematics, which together can substantially increase the experimental variance.  

If we assume that the density field is statistically isotropic, as it is in most well-motivated theoretical models, then
the bispectrum $ B(k_1,k_2,k_3)$ is defined by  
\begin{align}
\<\d_{\bk_1}\d_{\bk_2}\d_{\bk_3}\> &= (2\pi)^3 \d_D(\bk_1+\bk_2+\bk_3) B(k_1,k_2,k_3)\,,
\end{align}
where $ \d_D ({\bf k})$ is the three-dimensional Dirac $\delta$-function enforcing a triangle condition on the wavevectors ${\bf k}_i$, for which it is sufficient to use only the wavenumbers $k_i = |\bk_i|$.   
For simplicity, let us suppose we are only in a mildly nonlinear regime 
with good observational coverage over a modest redshift range, so that we can make 
the approximation that the covariance matrix is nearly diagonal 
$C^{-1}(\d^{obs}_{\bk}) \approx {\d^{obs}_{\bk}}/{P(k)}$.
With these replacements, the estimator (\ref{eq:bestimatorgen}) becomes
\begin{align}\label{eq:bestimator}
\curl{E} = \int \frac{d^3k_1}{(2\pi)^3} \frac{d^3k_2}{(2\pi)^3} \frac{d^3k_3}{(2\pi)^3} \frac{(2\pi)^3 \d_D(\bk_1+\bk_2+\bk_3)B(k_1,k_2,k_3)}{P(k_1)P(k_2)P(k_3)}\left[\d^{obs}_{\bk_1}\d^{obs}_{\bk_2}\d^{obs}_{\bk_3}  - 3\langle
\d^{sim}_{\bk_1}\d^{sim}_{\bk_2}\rangle \d^{obs}_{\bk_3}\right]\,,
\end{align}
where $\d^{sim}_\bk$ represents simulated data with the known inhomogeneous systematic effects included, while 
we also assume that shot noise is incorporated in the power spectrum $P+N \rightarrow \tilde P$, along with 
incomplete sample coverage (though we will drop the tilde).  We note that, although this 
galaxy estimator with a linear term (\ref{eq:bestimator}) has not been given in this form explicitly before, 
the bispectrum scaling and signal-to-noise ratios here and in what follows are consistent with the pioneering discussions in  refs.~\cite{9704075,0312286} (see also the analogous CMB bispectrum 
estimator discussed in ref.~\cite{Crem} and elsewhere).
In any case, this large-scale structure bispectrum estimator (\ref{eq:bestimator}) does not appear to be 
particularly useful because its brute force evaluation would require at least $\lmax^6$ operations for a single measurement (after 
imposing the triangle condition).  The problem is compounded by the many simulated realizations of the 
observational set-up which are required to obtain an accurate linear term in (\ref{eq:bestimator}).  In fact, if the
 theoretical bispectrum $ B(k_1,k_2,k_3)$ is computed numerically, then this is even more computationally intensive,  
 since it requires  many $N$-body simulations and bispectrum evaluations to achieve statistical precision.       

Nevertheless, let us now suppose that we have a large set of simulated non-Gaussian realisations $\d^{obs}_k$ generated 
with the same theoretical bispectrum $B(k_1,k_2,k_3)$  (and the same power spectrum $P(k)$).   If we take the expectation 
value of the estimator (\ref{eq:bestimator}) by summing over these realisations, then we find the average to be  
\eqa\label{eq:bestav}
\<\curl{E}\> & =&  \int \frac{d^3k_1}{(2\pi)^3} \frac{d^3k_2}{(2\pi)^3} \frac{d^3k_3}{(2\pi)^3} (2\pi)^6 \d_D^2(\bk_1+\bk_2+\bk_3)  \frac{B^2(k_1,k_2,k_3)}{P(k_1)P(k_2)P(k_3)} \cr
&= & \frac{V}{\pi}\int_{\curl{V}_B}  dk_1 dk_2 dk_3 \,\frac{ k_1 k_2 k_3 \,B^2(k_1,k_2,k_3)}{P(k_1)P(k_2)P(k_3)}\,,
\qea
where ${\curl{V}_B}$ is the tetrahedral region allowed by the triangle condition.   The averaged estimator (\ref{eq:bestav}) is
an important expression, so it is instructive for subsequent calculations
to outline the explicit steps that take us between these two lines.   First, the second 
Dirac $\delta$-function contributes only a volume factor $\delta({\bf 0}) = V/(2\pi)^3$.   Secondly, we complete the angular 
integration by expanding the integral form of the remaining $\delta$-function in spherical Bessel functions and harmonics,
\begin{align}\label{eq:deltaint}
\d_D(\bk) &= \frac{1}{(2 \pi)^3} \int d^3x e^{i \bk \cdot \bx},\\\label{eq:exptobessel}
 e^{i \bk \cdot \bx} &=4\pi \sum_{l m} i^l j_l(kx) Y_{lm}(\uk)Y^*_{lm}(\ux)\,.
\end{align}
Thirdly, each $\uk_i$ integration involves just a single spherical harmonic and contributes a factor 
$2\sqrt{\pi}\,\d_{l0}\,\d_{m0}$, so we end up with only a constant term from the Gaunt integral $G^{000}_{000} = 1/2\sqrt{\pi}$
(i.e.\ the integration over the three remaining $Y_{lm}({\bf x})$). 
Finally, the last integral over
the three Bessel functions $j_0(k_1x)j_0(k_2x)j_0(k_3x)$ yields $\pi/4 k_1k_2k_3$ and simultaneously imposes a triangle condition on $k_1,\,k_2,\,k_3$ which we denote by the restricted domain of integration ${\curl{V}_B}$.  

The estimator average (\ref{eq:bestav}) leads naturally to a weighted cross-correlator or inner product between two different  
bispectra $B_i(k_1,k_2,k_3)$ and $B_j(k_1,k_2,k_3)$, that is,
\begin{align}\label{eq:innerprod}
\mathcal{C}(B_i,B_j)=\frac{\langle B_i, \, B_j\rangle}{\sqrt{\langle B_i, \, B_i\rangle \langle B_j, \, B_j\rangle}}\,,
\end{align}
where   
 \begin{align}
\langle B_i, \, B_j\rangle ~\equiv~
\frac{V}{\pi}\int_{\curl{V}_B}  dk_1 dk_2 dk_3 \,\frac{ k_1 k_2 k_3 \,B_i(k_1,k_2,k_3)\,B_j(k_1,k_2,k_3)}{P(k_1)P(k_2)P(k_3)}\,.
 \end{align}
The estimator (\ref{eq:bestimator}) is thus proportional to the Fisher matrix of the bispectrum, $F_{ij} = \mathcal{C}(B_i,B_j) / 6\pi$ (see ref.~\cite{0312286}). 

The fiducial model for nonGaussianity is the $\fnl=1$ local model.  For the CMB, where the final CMB bispectrum $\Blll$ is linearly related to the primordial bispectrum $B_0(\kall)$, it is straightforward to define a normalisation which yields a universal $\Fnl$, representing the total integrated bispectrum for a particular theoretical model relative to that from the $\fnl=1$ local model (see ref.~\cite{FLS10}).  However, with bispectrum contributions from gravitational collapse and nonlinear bias arising even with Gaussian initial conditions, a universal normalisation is a more subtle issue which we will defer to section~\ref{sec:fnlestimation}.

Finally, we point out that the bispectrum estimator (\ref{eq:bestimatorgen}) can be applied in any three-dimensional 
physical context where we wish to test for a particular non-Gaussian model.   It can be applied at primordial times, with 
potential fluctuations (i.e. replacing $\d_\bk \rightarrow \Phi_\bk$), in the late-time linear regime on large scales 
where the density perturbation is simply related by a transfer function $\delta_{\bf k} = T(k,z) \,\Phi_k$ (as in the CMB), 
in the mildly non-linear regime where next-to-leading order corrections are known, or
deep in the nonlinear regime on small scales 
where we must rely  on $N$-body and hydrodynamic simulations.   However, for a useful implementation, we
must rewrite (\ref{eq:bestimatorgen}) in a separable form. 

\subsection{Separable mode expansions and bispectrum reconstruction}

The averaged estimator (\ref{eq:bestav}) gives a natural measure for defining separable mode functions 
\begin{align}\label{eq:Qexp}
Q_n(\kall) = {\textstyle \frac{1}{6}}[q_r(k_1)\,q_s(k_2)\, q_t(k_3) + 5 \hbox {perms}] \equiv q_{\{r}(k_1)\,q_s(k_2)\, q_{t\}}(k_3)\,,
\end{align}
 which we can use to decompose an arbitrary bispectrum (here, for convenience, 
the label $n$, denotes a linear ordering of the 3D products $n\leftrightarrow\{rst\}$).   
We choose to expand the bispectrum  $B(\kall)$ in its noise-weighted form (see ref.~\cite{FLS09}),
\begin{align}\label{eq:bseparable}
\frac{B(k_1,k_2,k_3)\, v(k_1)v(k_2)v(k_3)}{\sqrt{P(k_1) P(k_2) P(k_3)}} = \sum \aQn \Qn(\kall)\,,
\end{align}
where we have used the freedom to introduce a separable modification to the weight function $w(\kall) = 
k_1k_2k_3/v^2(k_1)v^2(k_2)v^2(k_3)$ in (\ref{eq:bestav}).  Series convergence usually 
can be improved with scale-invariance, suggesting 
the choice $v(k) = \sqrt{k}$.    The exact form of the one-dimensional basis functions $q_r(k)$ is not important, 
except that they should be bounded and well-behaved on the bispectrum domain ${\cal V}_B$.  Some $q_r(k)$ examples
which are orthogonal on ${\cal V}_B$ were given explicitly in ref.~\cite{FLS09}, 
analogues of Legendre polynomials $P_n(k)$.    

The product functions $\Qn$ are independent  but not necessarily orthogonal, so it is convenient 
from these to generate an orthonormal set of mode functions $\Rn$, such that, $\langle \Rn,\,\Rm\rangle =\delta_{nm} $ (achieved using Gram-Schmidt orthogonalisation 
with the inner product (\ref{eq:innerprod})).   We distinguish the expansion coefficients $\aQn$ and $\aRn$ 
by the superscripts for the separable `$Q$' and orthonormal `$R$' modes respectively; these are related to each 
other by a rotation involving the matrices $\langle \Qm,\,\Qn\rangle$ and $\langle \Qm,\,\Rn\rangle$(see ref.~\cite{FLS09}).    The orthonormal modes $\Rn$ are convenient for finding the expansion coefficients of an arbitrary bispectrum $B(\kall)$ from the inner product (\ref{eq:innerprod}) through $\aRn = \langle B,\,\Rn\rangle$ which are then rotated to the
more explicitly separable form $\aQn$.  Of course, there is some computational effort ${\cal O}(\nmax\times\lmax^3)$
to achieve this orthogonalisation and decomposition, but it is a modest initial computation which creates 
a framework for the subsequent data and error analysis. 
 
Now consider the effect of substituting the expansion (\ref{eq:bseparable}) into the bispectrum estimator (\ref{eq:bestimator}). 
It collapses to the simple summation
\begin{align}\label{eq:bestsum}
\curl{E} =\sum_n \aQn\, \bQn\,,
\end{align}
where the observed $\bQn$ coefficients are defined by 
\begin{align}
\bQn = \int d^3x \,M_r(\bx)\, M_s(\bx)\, M_t(\bx)\,,
\end{align}
with $M_r(\bx)$ the observed density perturbation multiplied in Fourier space with the mode functions $q_r(k)$, that is, 
\begin{align}\label{eq:bfiltered}
M_p(\bx) = \int d^3k \frac{\d^{obs}_{\bk}q_r(k)\,e^{i \bk \cdot \bx}}{\sqrt{k P(k)}}\,.
\end{align}
Including the linear term in (\ref{eq:bestimator}) to account for systematic inhomogeneous effects we have 
\begin{align}
\bQn = \int d^3x  \(\,M_r(\bx)\, M_s(\bx)\, M_t(\bx) -  [\langle M_r(\bx)\, M_s(\bx)\rangle M_t(\bx)+ \hbox{2 perms}]\)\,.
\end{align}
Furthermore, rotating to the orthonormal frame with $\Rn$, it is straightforward to demonstrate that 
the averaged observed coefficient will be $\aRn= \langle \bRn\rangle$, given a set of realizations with 
the bispectrum $B(\kall)$ in (\ref{eq:bseparable}).  Thus we can directly reconstruct the bispectrum from a single realization (with 
sufficient single-to-noise) using
\begin{align}\label{eq:breconstruct}
B(\kall) = \frac{\sqrt{P(k_1) P(k_2) P(k_3)}}{\sqrt{k_1k_2k_3}}\,\sum_n \bRn \, \Rn(\kall)\,.
\end{align}
This reconstruction yields the full bispectrum shape in a model independent manner.   One can also consider a model independent measure of the total integrated non-Gaussian signal, using Parseval's theorem in the orthonormal frame (see ref.~\cite{FLS10} for a discussion of the quantity $\bar \Fnl^2 = \sum_n\bRn{}^2$). However, the  bispectrum estimator (\ref{eq:bestsum}) provides an immediate means to determine the significance of an observation of a particular type of nonGaussianity with specific coefficients $\aQn$, e.g.\ by comparison with the $\bRn$ extracted from Gaussian simulations. We note that an initial implementation of the bispectrum reconstruction method (\ref{eq:breconstruct}) indicates its efficacy in recovering local nonGaussianity.

We emphasise that the  bispectrum reconstruction (\ref{eq:breconstruct}) provides an extremely efficient method for calculating 
the bispectrum from any given density field $\d_\bk$ with optimum noise weighting.    Moreover, these separable mode 
expansion methods have been thoroughly tested in a CMB context \cite{FLS10}.   In essence, the $\lmax ^6$ 
operations required with the original estimator (or for a direct bispectrum calculation such as that described in 
ref.~\cite{9704075}) have been reduced to a series of $\lmax^3$ integrations given by (\ref{eq:bfiltered}).   
Of course, the number of mode coefficients 
depends on the rate of convergence of the expansion (\ref{eq:bseparable}) which is usually remarkably rapid.
For the CMB, a comprehensive survey of most theoretical bispectra in the literature required only 30 eigenmodes
for an accurate description at WMAP resolution \cite{FLS10}.  Even for a separable bispectrum in the linear regime (i.e. a terminating sum), we shall explain the advantages of using the well-behaved mode expansion
(\ref{eq:bseparable}).    The form of the next-to-leading order corrections for 
large-scale structure show no obvious pathologies which would alter  this 
convergence significantly in the mildly nonlinear regime (see later), and substantial efficiencies will remain even in highly nonlinear contexts.    This reconstruction approach (\ref{eq:breconstruct}) is ideally suited for 
$N$-body simulations where the bispectrum can be predicted at high precision by efficiently extracting it from 
multiple realizations 
using both Gaussian and nonGaussian initial conditions (see later).    In an observational context, sparse sampling 
or poor survey strategies could reduce the effectiveness of the estimator (\ref{eq:bestimator}) in Fourier space, 
so care must be taken in large scale structure survey design to ensure good 
coverage so that higher order correlator measurements exploit these efficiencies.  

\section{Extension to the trispectrum and beyond}\label{sec:trispectrum}

\subsection{General trispectrum estimator}

In \cite{RSF10} we discussed general CMB estimators for the trispectrum, where 
the decomposition of a planar trispectrum (non-diagonal or single diagonal) is sufficient to study the majority of 
cases described in the literature. While this projection depends explicitly on five parameters (or four in the 
non-diagonal case), in order to study other probes of nonGaussianity, particularly for nonlinear large-scale 
structure, it may be necessary to consider the general trispectrum depending on the full six parameters. 
This is further motivated by  the study of the galaxy 
bispectrum, which may contain an enhanced contribution due to the trispectrum (see, e.g., ref.~\cite{09040497}). 
Clearly, then, we should also include a non-zero trispectrum to obtain non-Gaussian initial 
conditions suitable for a general bispectrum analysis using $N$-body codes.

The form of the general trispectrum estimator, for the connected part of a given four-point correlator $\<\d_{\bk_1}\d_{\bk_2}\d_{\bk_3}\d_{\bk_4}\>_c$, is directly analogous
to that presented already in ref.~ \cite{RSF10} for the CMB:
\begin{align}\label{eq:testimator}
\curl{E} = \int \frac{d^3\bk_1}{(2\pi)^3} \frac{d^3\bk_2}{(2\pi)^3} \frac{d^3\bk_3}{(2\pi)^3}\frac{d^3\bk_4}{(2\pi)^3} \frac{\(  \d^{\rm{obs}}_{\bk_1}\d^{\rm{obs}}_{\bk_2}\d^{\rm{obs}}_{\bk_3}\d^{\rm{obs}}_{\bk_4}-6\< \d^{\rm{sim}}_{\bk_1}\d^{\rm{sim}}_{\bk_2}\>\d^{\rm{obs}}_{\bk_3}\d^{\rm{obs}}_{\bk_4} +3\< \d^{\rm{sim}}_{\bk_1}\d^{\rm{sim}}_{\bk_2}\>\<\d^{\rm{sim}}_{\bk_3}\d^{\rm{sim}}_{\bk_4}\> \)\<\d_{\bk_1}\d_{\bk_2}\d_{\bk_3}\d_{\bk_4}\>_c}{P(k_1)P(k_2)P(k_3)P(k_4)},
\end{align}
where the notation $\<\dots \>_c$ denotes the connected component of the correlator. Note that this formula includes the quadratic term necessary to generalise to the case of incomplete sample coverage and inhomogeneous noise in a similar fashion to the CMB trispectrum estimator (see the discussion after (\ref{eq:bestimator})).  We omit the covariance-weighted version of the expression which is obvious from a comparison with (\ref{eq:bestimatorgen}).  
Imposing the $\d$-function appears to leave an intractable $\lmax^9$ operations for a full trispectrum estimator evaluation,
but, as with the bispectrum, this can be reduced dramatically using a separable approach. 

Assuming statistical isotropy, we can choose to parametrise the trispectrum using the lengths of four of its sides 
and two of its diagonals. In particular, we can exhibit these dependencies explicitly by representing the $\d$-function
imposing the quadrilateral condition, as a product of triangle conditions using the diagonals:
\begin{align}
 \<\d_{\bk_1}\d_{\bk_2}\d_{\bk_3}\d_{\bk_4}\>_c =& (2\pi)^3 \d_D(\bk_1+\bk_2+\bk_3+\bk_4) T(\bk_1,\bk_2,\bk_3,\bk_4)\\
=&(2\pi)^3\int d^3\mathbf{K}_1 d^3\mathbf{K}_2 \d_D(\bk_1+\bk_2-\mathbf{K}_1)\d_D(\bk_3+\bk_4+\mathbf{K}_1)\d_D(\bk_1+\bk_4-\mathbf{K}_2)T(k_1,k_2,k_3,k_4,K_1,K_2),
\end{align}
The decomposition of the trispectrum $T(k_1,k_2,k_3,k_4,K_1,K_2)$ is similar to that described in \cite{RSF10}, but in 
which the trispectrum is assumed to depend on the first five parameters only. In the interest of completeness we evaluate a suitable weight function necessary for
evaluation of the more general decomposition from the expectation value of the estimator (\ref{eq:testimator}).
Similarly to the case of the bispectrum (\ref{eq:bestav}), the expectation value for the estimator is found to take the 
following simple form:
\eqa\label{eq:testav}
\< \curl{E}\> &=& \frac{V}{(2\pi)^3}\int \frac{d^3\bk_1}{(2\pi)^3} \frac{d^3\bk_2}{(2\pi)^3} \frac{d^3\bk_3}{(2\pi)^3}\frac{d^3\bk_4}{(2\pi)^3} \frac{(2\pi)^6 \d_D(\bk_1+\bk_2+\bk_3+\bk_4) T^2(\bk_1,\bk_2,\bk_3,\bk_4)}{P(k_1)P(k_2)P(k_3)P(k_4)}\\\label{eq:TrispEstim}
&=& \frac{V}{(2\pi)^3}\frac{1}{2\pi^4}\int_{\curl{V}_T} d k_1 dk_2 dk_3 dk_4 dK_1 dK_2 \frac{k_1k_2 k_3 k_4 K_1 K_2}{\sqrt{g_1}}   \frac{T^2(k_1,k_2,k_3,k_4,K_1,K_2)}{P(k_1)P(k_2)P(k_3)P(k_4)},
\qea
where  the function $g_1$ is given by the expression  
\begin{align}\label{eq:gdef}
g_1=K_1^2 K_2^2 (\sum_i k_i^2 -K_1^2-K_2^2)-K_1^2 \kappa_{23}\kappa_{14}+K_2^2 \kappa_{1 2} \kappa_{3 4}-(k_1^2k_3^2-k_2^2k_4^2)(\kappa_{12}+\kappa_{34}),
 \end{align}
and we denote $\kappa_{ij}=k_i^2-k_j^2$.   For clarity, we omit the many calculational steps required in the derivation and 
present them in the Appendix.   Here, we note that $\curl{V}_T$ is the region allowed by the quadrilateral condition which is 
described in some detail in  \cite{RSF10}, noting the different ranges for the wavenumbers $k_i<k_{\rm{max}}$ and
diagonals $K_i<2 k_{\rm{max}}$.  By considering two different trispectra $T^2 \rightarrow T_iT_j$ in the 
estimator average (\ref{eq:testav}), we can use this expression
to define a noise-weighted cross-correlator and inner product (or Fisher matrix, see the discussion after (\ref{eq:bestav})).

\subsection{Separable mode expansions and the trispectrum estimator}

Using the weight \eqref{eq:testav}, a simple extension of the argument outlined in \cite{RSF10} to include two diagonals instead of one we find a similar eigenmode to the case of the bispectrum. In particular we could expand the trispectrum as $\w T(k_1,k_2,k_3,k_4,K_1,K_2)=\sum_n \alpha_n Q_n(k_1,k_2,k_3,k_4,K_1,K_2)$ where $Q_n=q_{\{r}(k_1)q_{s}(k_2)q_t(k_3) q_{u\}}(k_4) r_v(K_1) r_w(K_2)$, $n$ represents $\{r s t u v w\}$\footnote{The diagonals and the wavenumbers are described by different eigenmodes due to their differing range, i.e. $k_i<k_{\rm{max}}$ while $K_i<2 k_{\rm{max}}$. } and $\w$, here and subsequently, is shorthand for an appropriate separable weighting. As we will see in the estimator below, however,  it is simpler to achieve a separable form by parametrising our bispectrum using  angles rather than diagonals. To achieve this, we may make a coordinate transformation from $(K_1,K_2)\rightarrow (\mu=\hat{\bk}_1.\hat{\bk}_2,\nu=\hat{\bk}_1.\hat{\bk}_4)$ where we use $K_1=\sqrt{k_1^2+k_2^2+2k_1 k_2 \mu}$ and $K_2=\sqrt{k_1^2+k_4^2+2k_1 k_4\nu}$. The Jacobian of this transformation is $k_1^2 k_2 k_4/(K_1 K_2)$. Thus (\ref{eq:testav}) becomes
\begin{align}\label{eq:tinnerprod}
\< \curl{E}\> =& \frac{V}{(2\pi)^3}\frac{1}{2\pi^4}\int_{\curl{V}_T} d k_1 dk_2 dk_3 dk_4 d\mu d\nu \frac{k_1^3 k_2^2 k_3 k_4^2}{\sqrt{g_1}}   \frac{T^2(k_1,k_2,k_3,k_4,\mu,\nu)}{P(k_1)P(k_2)P(k_3)P(k_4)}.
\end{align}
where $g_1$ is given by equation~\eqref{eq:gdef} but now must be expressed in terms of $\mu,\nu$. We may use this weight to form an eigenmode expansion of the trispectrum where we use Legendre polynomials to describe the angular part. Explicitly we may expand the trispectrum in noise-weighted form as
\begin{align}\label{eq:tdecomp}
\frac{v(k_1)v(k_2)v(k_3)v(k_4)}{\sqrt{P(k_1)P(k_2)P(k_3)P(k_4)}}T(k_1,k_2,k_3,k_4,\mu,\nu)=\sum_{n l_1 l_2} \alpha_{n l_1 l_2} Q_n(k_1,k_2,k_3,k_4)P_{l_1}(\mu)P_{l_2}(\nu)
\end{align}
where $n=\{r,s,t,u\}$ and $Q_n(k_1,k_2,k_3,k_4)= q_{\{r} (k_1) q_s(k_2)  q_t (k_3) q_{u\}}(k_4)$ in an analogous manner to equation \eqref{eq:Qexp}. Scale invariance suggests the choice $v(k)=k^{3/4}$. In order to make this expression separable in terms of the vectors $\bk_i$ we note the following expansion of the Legendre polynomials
\begin{align}\label{Legendre}
 P_l(\hat{\bk}_1.\hat{\bk}_2)=\frac{4\pi}{2 l+1}\sum_{m=-l}^{l} Y_{l m}(\hat{\bk}_1)Y_{l m}^*(\hat{\bk}_2).
\end{align}
Using equations \eqref{eq:deltaint} and \eqref{eq:exptobessel} we can now write the estimator as expressed in \eqref{eq:testimator} in the form 
\begin{align}\label{eq:testsum}
\curl{E} = \sum_{n l_1 l_2} \baQ_{n l_1 l_2}\bbQ_{n l_1 l_2}\,,
\end{align} 
where the extracted trispectrum coefficients are given by 
\begin{align}
\bbQ_{n l_1 l_2}=& \frac{(4\pi)^2}{(2l_1+1)(2l_2+1)}\sum_{m_1 m_2}  \int d^3\bx  \Bigg[ M_{r l_1 l_2}^{m_1 m_2}(\bx)M_{s l_1}^{m_1*}(\bx)M_{t}(\bx)M_{u l_2}^{m_2 *}(\bx)\nonumber\\
&-\left(M_{r l_1 l_2}^{m_1 m_2}(\bx)M_{s l_1}^{m_1*}(\bx)\langle M_{t}(\bx)M_{u l_2}^{m_2 *}(\bx)\rangle+\rm{5\,perms}\right)
+\left(\langle M_{r l_1 l_2}^{m_1 m_2}(\bx)M_{s l_1}^{m_1*}(\bx)\rangle \langle M_{t}(\bx)M_{u l_2}^{m_2 *}(\bx)\rangle+\rm{2\,perms}\right)\Bigg],
\end{align}
where the permutations are with respect to the indices $\{r,s,t,u\}$.
In the above we define the filtered density perturbations $M^{...}_{...}$ by 
\begin{align}\label{eq:mapstrisp}
M_{r l_1 l_2}^{m_1 m_2}(\bx)&=\int \frac{d^3 \bk}{(2\pi)^3} e^{i\bk.\bx} \frac{q_r(k)\d_{\bk}^{\rm{obs}}}{\sqrt{P(k)}k^{3/4}} Y_{l_1 m_1}(\hat{\bk})Y_{l_2 m_2}(\hat{\bk}),\qquad
M_{s l_1}^{m_1 *}(\bx)=\int \frac{d^3 \bk}{(2\pi)^3} e^{i\bk.\bx} \frac{q_s(k)\d_{\bk}^{\rm{obs}}}{\sqrt{P(k)}k^{3/4}}Y_{l_1 m_1}^*(\hat{\bk}),\nonumber\\
M_t (\bx)&= \int \frac{d^3 \bk}{(2\pi)^3} e^{i\bk.\bx} \frac{q_t(k)\d_{\bk}^{\rm{obs}}}{\sqrt{P(k)}k^{3/4}},
\end{align}
with a  $*$ denoting a filtered map using $Y_{l m}^*$.

The algorithm (\ref{eq:testsum}) provides a highly efficient method for  estimating any trispectrum from a given 
 density field.  It requires only ${\cal O}(\nmax^{4/3}\times 
\lmax^3)$ operations, which makes feasible the intractable naive brute force calculation 
requiring ${\cal O}(\lmax^9)$ operations.  
In making this rough numerical estimate, we assume that the number of modes in each 
of the six dimensions is equal (and small), while noting that we have to perform a double summation for the two 
angle parameters $\mu,\,\nu$ over the indices $l_1,\,m_1,\,l_2,\,m_2$.   

As for the bispectrum, it is possible from the separable $\barQ_{nl_1l_2}$ modes to create a set of orthonormal 
 $\barR_{nl_1l_2}$ modes using the inner product (\ref{eq:tinnerprod}). 
Like the original decomposition of a theoretical trispectrum (\ref{eq:tdecomp}), 
orthogonalisation is a computationally intensive task
requiring up to ${\cal O}(\lmax^6)$ operations.  However, it  need only be performed once at the outset to set up 
the calculation framework, with the 
resulting rotation matrices being available for all the repetitive  subsequent analysis ($\sim \lmax^3$
operations).   We can realistically envisage, then, reconstructing the complete trispectrum directly from the observational data using the rotated $\bbQ_{n l_1 l_2}$ coefficients (as in  (\ref{eq:breconstruct}).   It is interesting to note 
that almost all theoretical trispectra presented to date in the literature are `planar', that is, either depending on only 
one diagonal or none.  We treat the latter special case below, but we leave the simplifications
arising from the single
diagonal case for discussion elsewhere \cite{Regan2011}.

\subsection{Non-diagonal trispectrum and quadspectrum estimation}

In the case that the trispectrum is independent of the diagonals $K_1,K_2$ (or angles $\mu$, $\nu$) we get a simpler 
expression for the averaged estimator (\ref{eq:testimator}):
\begin{align}
\langle \mathcal{E}\rangle=\frac{V}{ (2\pi)^6}\int_{\curl{V}_T} d k_1 dk_2 dk_3 dk_4 k_1 k_2 k_3 k_4 \Big(\sum_i k_i-|\tilde{k}_{34}|-|\tilde{k}_{24}|-|\tilde{k}_{23}|\Big)\frac{T^2(k_1,k_2,k_3,k_4)}{P(k_1)P(k_2)P(k_3)P(k_4)}
\end{align}
where $\tilde{k}_{34}=k_1+k_2-k_3-k_4$, etc. We may use the weighting this suggests to decompose the trispectrum into the form $\w T=\sum_n \alpha_n Q_n$ where $Q_n=q_{\{r} q_s q_t q_{u\}}$. The estimator is simpler to calculate since there are no cross terms between integrals. We find the extracted observational coefficients simplify to  
\begin{align}\label{eq:tndest}
 \beta_n=&\int d^3\bx \Bigg[ M_r (\bx) M_s (\bx) M_t(\bx) M_u(\bx)-\left( M_r (\bx) M_s (\bx) \langle M_t(\bx) M_u(\bx)\rangle +\rm{5\,perms}\right)\nonumber\\
 & +\left(\langle M_r (\bx) M_s (\bx)\rangle \langle M_t(\bx) M_u(\bx)\rangle +\rm{2\,perms}\right) \Bigg],
\end{align}
where $M_t$ was defined in \eqref{eq:mapstrisp}.   Here, we see that the trispectrum estimation scales once 
again as only ${\cal O}(\nmax\times\lmax^3)$ operations.  The extraction of expansion coefficients $\baQ$ 
from a given non-separable theoretical trispectrum appears to require up to $\lmax^4$ operations, but it is a one-off
calculation amenable to many shortcuts.   A practical implementation reveals that non-diagonal 
trispectra given in the literature require only $\nmax \approx {\cal O}(10)$ modes for accurate representation. 
As an example, even the pathological local model with diverging squeezed states requires only $\nmax =20$ for the 
expansion (\ref{eq:tdecomp}) to achieve a 95\% correlation with the primordial shape.   It is clear that there is 
no inherent impediment to direct  estimation and evaluation of trispectra from survey data of adequate quality.  

This separable methodology can be applied to correlators beyond the trispectrum, such as the quadspectrum
$\tilde {\cal Q}(\bk_1,\bk_2,\bk_3,\bk_4,\bk_5)$ defined from
\begin{align}
\langle \d_{\bk_1} \d_{\bk_2} \d_{\bk_3} \d_{\bk_4} \d_{\bk_5}\rangle=(2\pi)^3\d(\bk_1+\bk_2+\bk_3+\bk_4+\bk_5)\tilde{\cal Q}(\bk_1,\bk_2,\bk_3,\bk_4,\bk_5)\,.
\end{align}
For simplicity, however, we restrict attention here to quadspectra that are non-diagonal, depending only on the wavenumbers $k_1,\dots,k_5$, that is,  $\tilde{\cal Q}(\bk_1,\bk_2,\bk_3,\bk_4,\bk_5)=\tilde{\cal Q}(k_1,k_2,k_3,k_4,k_5)$. 
The expectation value of the quadspectrum estimator is then given by 
\begin{align}\label{eq:qinnerprod}
\langle\mathcal{E}\rangle&=\frac{V}{(2\pi)^3}\int \left(\Pi_{i=1}^5 \frac{d^3 \bk_i}{(2\pi)^3}\right)\frac{(2\pi)^6 \d(\bk_1+\bk_2+\bk_3+\bk_4+\bk_5)\tilde{Q}^2(k_1,k_2,k_3,k_4,k_5)}{P(k_1)P(k_2)P(k_3)P(k_4)P(k_5)}\nonumber\\
&=\frac{V}{(2\pi^3)^3}\int dk_1dk_2 dk_3 dk_4 dk_5 (k_1 k_2 k_3 k_4 k_5)^2\left( \int dx x^2 j_0 (k_1 x)  j_0 (k_2 x) j_0 (k_3 x) j_0 (k_4 x) j_0 (k_5 x)\right)\nonumber\\
&\qquad\qquad\qquad \times\frac{\tilde{Q}^2(k_1,k_2,k_3,k_4,k_5)}{P(k_1)P(k_2)P(k_3)P(k_4)P(k_5)}\,,
\end{align}
where the integral over the five spherical Bessel functions serves also to define the allowed quadspectrum domain 
${\cal V}_Q$.   
The expression (\ref{eq:qinnerprod}) may be used to derive a weight to decompose the quadspectrum in the form $\Big[\Pi_{i=1}^5 v(k_i)/\sqrt{P(k_i)}\Big]\tilde{\cal Q}(k_1,k_2,k_3,k_4,k_5)=\sum_n \alpha_n Q_n(k_1,k_2,k_3,k_4,k_5)$ where $n\leftrightarrow \{r,s,t,u,v\}$ and  $Q_n(k_1,k_2,k_3,k_4,k_5)=q_{\{r}(k_1)q_s(k_2)q_t(k_3)q_u(k_4)q_{v\}}(k_5)$, and where imposing scale invariance sets $v(k)=k^{9/10}$.   The resulting separable estimator is directly analogous to that for the non-diagonal trispectrum (\ref{eq:tndest}), but for brevity we will only discuss initial
conditions with a non-trivial quadspectrum.

\section{Efficient generation of arbitrary non-Gaussian initial conditions}\label{sec:intcon}

The generation of non-Gaussian initial conditions for $N$-body simulations with a given primordial bispectrum has been 
achieved to date only for bispectra which have a simple separable form (see, e.g., \cite{0701131,07072516,07104560,
10065793}).  For $N$-body codes to efficiently produce non-Gaussian initial conditions for an arbitrary non-separable bispectrum, will require a well-behaved separable mode decomposition, as achieved for CMB map simulations 
 in ref.~\cite{FLS09}.   However, we can do even better by simulating initial data given both an arbitrary 
 bispectrum and trispectrum, as shown for the CMB in \cite{RSF10}.  As we have discussed already, 
 this is of particular interest for measurements of the large-scale structure bispectrum, because of nonlinear contributions
expected from the trispectrum.   We describe the non-Gaussian primordial potential perturbation as 
\begin{align}\label{eq:intcondexp}
\O = \O^G +\frac{1}{2} \Fnl \O^B +\frac{1}{6} G_{NL} \O^T,
\end{align}
where $\O^G$ is a Gaussian random field with the required power spectrum $P(k)$. It should be noted that this definition introduces two trispectrum terms of the form $\langle \O^T \O^G \O^G \O^G \rangle$ and $\langle \O^B \O^B \O^G \O^G \rangle$ (similar to the local trispectrum terms with coefficients $g_{NL}$ and $\tau_{NL}$ respectively). Therefore, it may be desirable to cancel this extra contribution. This issue will be addressed at the end of the section. Following ref.~\cite{FLS09} for 
the primordial bispectrum $B(\kall)$ with separable expansion
\begin{align}\label{symmexpan}
\frac{B(k,k^{'},k^{''}) }{P(k^{'}) P(k^{''})+P(k^{}) P(k^{'})+P(k^{}) P(k^{''})}=\sum_{n}\alpha_n^Q Q_n(k,k',k''),
\end{align}
the  bispectrum contribution to the primordial perturbation $\O$ becomes simply
\begin{align}\label{eq:icbi}
\O^B(\bk) &= \int \frac{d^3\bk^{'}}{(2\pi)^3}\frac{d^3\bk^{''}}{(2\pi)^3} \frac{(2\pi)^3 \d(\bk+\bk^{'}+\bk^{''}) B(k,k^{'},k^{''}) \O^G(\bk^{'}) \O^G(\bk^{''})}{P(k^{'}) P(k^{''})+P(k^{}) P(k^{'})+P(k^{}) P(k^{''})}\,,\\
=&\sum_{n}\alpha_{n} q_{\{r}(k)\int d^3 \bx e^{i\bk.\bx}M_s(\bx)M_{t\}}(\bx),
\end{align}
where the filtered density perturbations $M_s(\bx)$ are now defined by
\begin{align}\label{eq:b2filtered}
M_s(\bx) = \int \frac{d^3\bk}{(2\pi)^3} \Phi^{G}({\bk}) q_s(k)\,e^{i \bk \cdot \bx}\,.
\end{align}
We note that the modal bispectrum algorithm in ref.~\cite{FLS09} used here is a generalization of the 
separable CMB bispectrum simulation method presented in ref.~\cite{0612571}.   Here, in 3D, 
the intermediate expression in (\ref{eq:icbi} was first presented in convolved form (see (\ref{eq:convolved}) below) in 
refs.~\cite{10065793,VerdeWag}.
It should be noted that, with this prescription, the definition agrees identically with the expansion $\O=\O^G+\Fnl \O^G*\O^G$ in the case of the local model.  Of course, we normalise $B(\kall)$ such that it has $\Fnl =1$.
Like the estimator, this requires only ${\cal O}(\nmax\times\lmax^3)$ operations for every realization of new initial conditions, as opposed to a brute force approach which requires $\lmax^6$.  Note also that once the $\nmax$ filtered density perturbations $\int d^3 \bx e^{i\bk.\bx}M_s(\bx)M_{t\}}(\bx)$ have been obtained for a given $\O^G$, they can be applied to an arbitrary number of different shaped bispectra represented by $\aQn$s.

We can similarly find a relatively simple and highly efficient expression to compute initial conditions for the 
trispectrum $\Phi^T$. 
Following \cite{RSF10},
the primordial trispectrum $T(\kall,\,k_4,\,\mu,\,\nu)$ is represented and expanded using wavenumber $q_r(k)$ and angle $P_u(\mu)$ modes in a similar fashion to equation (\ref{eq:tdecomp}), 
\begin{align}\label{eq:tdecomp2}
\frac{ T(k_1,k_2,k_3,k_4,\mu,\nu)}{{P(k_1)P(k_2)P(k_3)P(k_4)}+3\,{\mathrm{perms}}}=\sum_{n l_1 l_2} \alpha_{n l_1 l_2} Q_n(k_1,k_2,k_3,k_4)P_{l_1}(\mu)P_{l_2}(\nu).
\end{align}

The trispectrum contribution to $\O$ then becomes
\begin{align}\label{eq:ictri}
\Phi^T(\bk)=& \int \frac{d^3\bk^{'}d^3\bk^{''}d^3\bk^{'''}}{(2\pi)^6} \frac{ \d(\bk+\bk^{'}+\bk^{''}+\bk^{'''}) T(\bk,\bk^{'},\bk^{''},\bk^{'''}) \O^G(\bk^{'}) \O^G(\bk^{''})\O^G(\bk^{'''})}{P(k^{'}) P(k^{''}) P(k^{'''})+3\,{\mathrm{perms}}}
\\
=&\sum_{n l_1 l_2}\baQ_{n l_1 l_2}\frac{(4\pi)^2}{(2l_1+1)(2l_2+1)}\sum_{m_1 m_2}Y_{l_1 m_1}(\hat{\bk})Y_{l_2 m_2}(\hat{\bk}) q_r(k)\nonumber\\
&\times\int d^3 \bx e^{i\bk.\bx}M_{s l_1}^{m_1*}(\bx)M_t(\bx)M_{u l_2}^{m_2*}(\bx),
\end{align}
where the filtered density perturbations $M_{s l_1}^{m_1*}$ and $M_t$ are now given by
\begin{align}\label{eq:mapstrisp}
M_{s l_1}^{m_1 *}(\bx)&=\int \frac{d^3 \bk}{(2\pi)^3} e^{i\bk.\bx} {q_s(k)\O^G({\bk})}Y_{l_1 m_1}^*(\hat{\bk}),\qquad \nonumber\\
M_t (\bx)&= \int \frac{d^3 \bk}{(2\pi)^3} e^{i\bk.\bx} q_t(k)\O^G({\bk}).
\end{align}

For the particular case that the trispectrum is independent of the angles $\mu,\,\nu$ (or diagonals $K_1,\,K_2$) the decomposition is somewhat simpler:
\begin{align}
 \Phi^T(\bk) =\sum_n \baQ_n q_r(k)\int d^3\bx e^{i\bk.\bx} M_s (\bx) M_t(\bx) M_u(\bx)\,.
\end{align}
This applies to many cases in the literature, including constant, local and equilateral models.   
This simplification 
will also apply to initial conditions with non-diagonal quadspectra.
The expression for quadspectrum perturbation $\Phi^{\tilde{Q}}$ is very similar to the expressions above with 
\begin{align}
\Phi^{\tilde{Q}}=\sum_n \tilde\alpha^{\scriptstyle Q}_n \,q_r(k) \int d^3 \bx e^{i\bk.\bx}M_s(\bx)M_t(\bx)M_u(\bx)M_v(\bx).
\end{align}
It is clear that it is possible, given separable expansions of an arbitrary bispectrum and trispectrum, to efficiently 
generate multitudes of realizations, with each requiring only ${\cal O}(\nmax\times\lmax^3)$ operations.

It should be noted that the since
that the bispectrum (\ref{eq:icbi}) and trispectrum (\ref{eq:ictri}) contributions are not independent, it may be necessary to subtract out an unwanted `bispectrum' contribution to the trispectrum. The bispectrum contribution induces a trispectrum given by
\begin{align}
\langle\O(\bk_1)\O(\bk_2)\O(\bk_3)\O(\bk_4)\rangle_c & \nonumber\\
=(2\pi)^3 F_{NL}^2 \int d^3\bK& \Big[\tilde{T}(k_1,k_2,k_3,k_4,K)\delta_D(\bk_1+\bk_2-\bK)\delta_D (\bk_3+\bk_4+\bK)\nonumber\\
&+\tilde{T}(k_1,k_3,k_2,k_4,K)\delta_D(\bk_1+\bk_3-\bK)\delta_D (\bk_2+\bk_4+\bK)\nonumber\\
&+\tilde{T}(k_1,k_4,k_2,k_3,K)\delta_D(\bk_1+\bk_4-\bK)\delta_D (\bk_2+\bk_3+\bK)\Big],
\end{align}
where 
\begin{align}
\tilde{T}(k_1,k_2,k_3,k_4,K)=&\frac{B(k_1,k_2,K)}{P(k_1)P(k_2)+2\,{\mathrm{perms}}} \frac{B(k_3,k_4,K)}{P(k_3)P(k_4)+2\,{\mathrm{perms}}}P(K)\nonumber\\
&\times\left(P(k_1)P(k_3)+P(k_1)P(k_4)+P(k_2)P(k_3)+P(k_2)P(k_4)\right).
\end{align}
Cancellation of this spurious `trispectrum' may be achieved by altering the algorithm given by equation \eqref{eq:intcondexp} to the form
\begin{align}\label{eq:intcondexp2}
\O = \O^G +\frac{1}{2} \Fnl \O^B +\frac{1}{6} G_{NL} \O^T-\frac{1}{2} \Fnl^2 \tilde{\O}^T,
\end{align}
where
\begin{align}
\tilde{\O}^T(\bk)=&\int \frac{d^3 \bk_2}{(2\pi)^3} \frac{d^3 \bk_3}{(2\pi)^3} \frac{d^3 \bk_4}{(2\pi)^3}d^3 \bK(2\pi)^3 \delta_D(\bk+\bk_2-\bK) \delta_D(\bk_3+\bk_4+\bK)\nonumber\\
&\times\frac{\tilde{T}(k,k_2,k_3,k_4,K)}{P(k)P(k_2)P(k_3)+3\,{\mathrm{perms}}}\O^{G}(\bk_2)\O^{G}(\bk_3)\O^{G}(\bk_4).
\end{align}
With this prescription it is found that 
\begin{align}
\langle \O(\bk_1)\O(\bk_2)\O(\bk_3) \rangle&=(2\pi)^3 \delta_D(\bk_1+\bk_2+\bk_3)B(k_1,k_2,k_3)\, ,\nonumber\\
\langle \O(\bk_1)\O(\bk_2)\O(\bk_3) \O(\bk_4)\rangle_c&=(2\pi)^3 \delta_D(\bk_1+\bk_2+\bk_3+\bk_4)T(\bk_1,\bk_2,\bk_3,\bk_4),
\end{align}
as desired. We shall leave a detailed analysis of this issue to a future work.

Recently, refs.~\cite{10065793,VerdeWag} proposed an alternative approach to creating non-Gaussian 
initial conditions from bispectra by integrating directly the convolution expression
\begin{align}\label{eq:convolved}
\O^B(\bk) &= \int \frac{d^3\bk^{'}}{(2\pi)^3}\frac{B(k,k^{'},|\bk+\bk^{'}|) \, \O^G(\bk^{'}) \, \O^G(\bk+\bk^{'})}{P(k^{'}) P(|\bk+\bk^{'}|)+P(k^{}) P(k^{'})+P(k^{}) P(|\bk+\bk^{'}|)}\,.
\end{align}
Originally in ref.~\cite{10065793} the denominator only had a $P(k^{'}) P(|\bk+\bk^{'}|)$ term, so for explicitly separable bispectra,  using convolutions, they were able to exploit the same efficiencies 
described above to reduce the problem from ${\cal O}(\lmax^6)$ to ${\cal O}(\lmax^3)$ operations. However, 
this procedure leads in general to a non-trivial and spurious  non-Gaussian contribution to the power spectrum, 
so the above expression with a symmetrised denominator was advocated instead \cite{VerdeWag}.   
The key difficulty with this modification, however, is that the denominator becomes non-separable, so the method can no longer 
exploit separability in evaluating the convolution (except in the trivial local case where the integrand is unity).   
For models other than local, a highly inefficient brute force analysis was pursued.   We contrast this with the modal approach where
the problem of separable efficiency is already solved in general.   The modal decomposition does not require the bispectrum 
$\Bkkk$ to be separable, 
so the form of the denominator in (\ref{symmexpan}) presents no additional difficulty.   In addition, we note that even in the 
for separable bispectra, the CMB modal initial conditions prescription had other beneficial effects because of the well-behaved bounded 
mode functions employed; these may carry over to this three-dimensional case.

\section{Non-Gaussian parameter estimation}\label{sec:fnlestimation}

Fast separable methods for estimating arbitrary bispectra or trispectra
in large scale structure observations or simulated data greatly improve 
the prospect of using higher order correlators as an important cosmological
 diagnostic.  This is particularly pertinent for testing the Gaussian hypothesis of 
 the inflationary scenario.   The complication is that even Gaussian initial 
fluctuations receive non-Gaussian contributions through late-time gravitational collapse (see reviews \cite{LigSef2010,10035020} and the references therein). 
Here, we briefly sketch some key issues facing parameter estimation in this context. 

There has been much recent progress describing next-to-leading order contributions to nonGaussianity
from gravity.   A simple example of this is the matter density power spectrum which contains
several contributions, including those from an enhanced primordial bispectrum $\Fnl B_0(\kall)$ \cite{08084085}:
\begin{align}\label{eq:PNG}
P^B(k)=\frac{\Fnl}{(2\pi)^3}\int d^3 \by B_0(\bk,\by,\bk-\by)F_2(\by,\bk-\by)=\frac{\Fnl}{(2\pi)^3}\int d^3 \by d^3 \bk_2 \d(\bk_2-\bk+\by) B_0(k,y,k_2)F_2(\by,\bk_2),
\end{align}
where the gravitational kernel for this convolution is given by 
\begin{align}\label{eq:F2exp}
F_2(\by,\bk_2)&=\frac{17}{21}+P_1(\mu)\left(\frac{y}{k_2}+\frac{k_2}{y}\right)+\frac{4}{21}P_2(\mu)\,.
\end{align}
Taking the separable expansion (\ref{eq:bseparable}) for $B_0(\kall)$ and substituting into eqn (\ref{eq:PNG}), we
find the simple integral over the mode functions $q_r(k)$: 
\begin{align}\label{eq:PB}
P^B(k)=&\Fnl\sum_n \frac{\alpha_n}{2\pi^2} \frac{q_r(k)\sqrt{P(k)}}{k^{3/2}}\int_{\mathcal{V}_B} dy dk_2 \, \sqrt{yP(y)} \, q_s(y) \,\sqrt{k_2P(k_2)} \,q_t(k_2)\nonumber\\
&\times\Big[ \frac{5}{7}+\frac{2}{7}\left(\frac{k_2^2+y^2-k^2}{2 k_2 y}\right)^2-\left(\frac{y}{k_2}+\frac{k_2}{y}\right)\left(\frac{k_2^2+y^2-k^2}{2 k_2 y}\right)\Big],
\end{align}
where $\mathcal{V}_B$ represents the domain for which the triangle condition holds for the wavenumbers $(k_2,y,k)$. 
Note that this integral breaks down into products of one dimensional integrals over $y$ and $k_2$ which can be 
evaluated easily.   Here, the calculation steps leading to (\ref{eq:PB}) are very similar to those used to obtain (\ref{eq:bestav}).

In the mildly nonlinear regime, the matter density bispectrum similarly 
contains nonlinear contributions  from gravitational collapse, from the 
primordial bispectrum $\Fnl B_0$,  and from the primordial trispectrum $\t_{\rm NL}T_0$ \cite{09040497,09050717}:
\begin{align}\label{eq:BT}
B(k_1,k_2,k_3)&=[2 F_2(\bk_1,\bk_2 P_0(k_1) P_0(k_2) + \hbox{2 perms}]+   \Fnl B_0(\kall) ]\\
&~~~+\frac{\t_{\rm NL}}{(2\pi)^3}\int d^3 \by  T_0(\bk_1,\bk_2,\by,\bk_3-\by)F_2(\by,\bk_3-\by)+\hbox{2 perms}\,.\nonumber\\
&\equiv B^G(\kall) + \Fnl B_0(\kall) +\t_{NL}B^T(\kall)\nonumber
\end{align}
In  Appendix B, we substitute the separable expansion for the trispectrum (\ref{eq:tdecomp}) into (\ref{eq:BT}) to 
find integral expressions for the resulting bispectrum.  For non-diagonal trispectra, the result is simple and 
very similar to the power spectrum modification (\ref{eq:PB}).   The result is three distinct contributions 
to the late-time bispectrum $\w B(\kall) = \sum_n\alpha_nQ_n$ with the bispectrum approximated as in 
separable form as 
\begin{align}
\w B(\kall)  = \sum_n (\alpha^G_n +\Fnl \alpha^B_n +\t_{NL} \alpha^T_n)\,\Rn(\kall)\,,
\end{align}
with the coefficients $\alpha^i_n$ representing distinct shapes in the orthonormal frame.   Here, the primordial $\alpha^B$ coefficients are normalised
such that in the initial conditions $\Fnl = 1$, and similarly for the primordial trispectrum $\t_{\rm NL}=1$. 

Setting aside the trispectrum contribution, if can remove the Gaussian part from $\alpha_n,\,\beta_n$ 
then we have an optimal estimator for the nonGaussianity parameter $\Fnl$ ,
\begin{align}\label{eq:estimatorfnl}
\curl{E} = \frac{1}{{N}^2}\sum \a^{B}_n \b^{B}_n\,,
\end{align}
where we have defined the predicted $\a^{B}_n$ and measured $\b^{B}_n$ by
\begin{align}
\a^{B}_n = \a_n - \bar{\a}^G_n\,, \qquad
\b^{B}_n = \b_n - \bar{\b}^G_n\,, \qquad
{N}^2 =  \sum {\a^{B}_n}^2\,.
\end{align}
Here $\bar{\a}^G_n$ refers to the decomposition coefficients for Gaussian initial conditions, calculated either from theory (as above in (\ref{eq:BT})) or obtained from $N$-body simulations (note $\bar{\a}^G_n=\bar{\b}^G_n$)  and the $\a_n$  are calculated from initial conditions with $\Fnl=1$. The variance of the estimator can then be calculated by applying it to a large set of Gaussian simulations. This is directly analogous to the CMB estimator used in \cite{FLS09} (where of course $\bar{\a}^G_n = 0$).  

However, in the nonlinear regime, and with significant bias affecting the galaxy distribution, it will not be possible to  approximate nonGaussianity in this simple way. We need to approach parameter estimation for $F^{NL}$ (or $\t_{\rm NL}$) quite differently. The estimator (\ref{eq:estimatorfnl}) can be thought of as a least squares fit of the theory to the data. As the relative size of the individual $\a^{B}_n$ are constant, we can only change the amplitude, $\Fnl$, we must simply choose a $\Fnl$ which minimises
\begin{align}
\curl{E} = \sum \( \a^{B}_n \Fnl - \b^{B}_n \)^2
\end{align}
for a given form of $\a^{B}_n$. In the general case we expect the ratios of the individual coefficients to change as we change $\Fnl$. As a result we must consider the $\a_n$ to be an arbitrary function of $\Fnl$ and so we now wish to minimise
\begin{align}
\curl{E}(\Fnl) = \sum \( \a_n(\Fnl) - \b_n \)^2
\end{align}
with respect to $\Fnl$. We will assume that it will not be possible in general to determine $\a_n(\Fnl)$ analytically so that we could then try to solve $\p \curl{E} / \p \Fnl = 0$. This means that to minimise $\curl{E}$ requires extracting the $\a_n$ from sets of $N$-body simulations each with different non-Gaussian initial conditions which correspond to a particular $\Fnl$. We then reconstruct the dependence of $\curl{E}$ on $\Fnl$ and find the best-fit $\Fnl$ for the
given observations.
One also must be careful calculating the variance on such a measurement of $\Fnl$. In general this would entail applying the same approach to each density distribution in the set of simulations with the estimated $\Fnl$ and then determining the distribution of the recovered $\Fnl$.  Of course, Gaussian simulations may be substituted if $\Fnl$ is sufficiently small that the effect on the error bars is negligible.  

Finally, we note that in general the galaxy bispectrum will take contributions from both the bispectrum and trispectrum of the curvature perturbation \cite{09040497} (which is why we cannot in general connect $\Fnl$ with its CMB counterpart in a simple way).  The amplitudes  of $\Fnl$ and $\t_{\rm NL}$ can be determined by consistency conditions for certain models  or they can vary independently. In this case we must constrain the amplitude of both $\Fnl$ and $\t_{\rm NL}$ contributions marginalising over these two parameters. Such a computationally intensive analysis becomes much more feasible 
with an efficient bispectrum extraction method (\ref{eq:breconstruct}) and with 
non-Gaussian initial conditions which include the specification of the trispectrum (\ref{eq:intcondexp}).

\section{Conclusion}
While the CMB is an ideal observable for tests of primordial nonGaussianity since the perturbations remain in the linear regime, the prospects for achieving comparable, and ultimately superior, constraints on nonGaussianity in the near future using large-scale structure appears encouraging due to recent advancements in the analysis and development of N-body codes.

In this paper we have described how methods developed for the analysis of nonGaussianity in the CMB may be applied to surveys of large-scale structure. These methods are based on mode expansions, exploiting a complete orthonormal eigenmode basis to efficiently decompose arbitrary poly-spectra into a separable polynomial expansion.

Applying the methodology to the bispectrum reveals a vast improvement in computational speed for finding a general estimator and correlator, reducing complexity from $\mathcal{O}(l_{\rm{max}}^6)$ to $\mathcal{O}(n_{\rm{max}}\times l_{\rm{max}}^3)$. As we use a complete orthonormal basis we are also able to efficently calculate the bispectrum from simulations and, assuming sufficent signal to noise, observations. Of particular interest is the application to the generation of nonGaussian initial conditions for N-body codes. The approach can be used to create initial conditions with arbitrary independent poly-spectra. With this method calculation of the bispectrum contribution requires a similar number of operations as decomposition. This improvement to the brute force approach opens up the opportunity of investigating a far wider range of models using large-scale structure than has hitherto been considered.

The extension of the approach to the trispectrum has also been described in some detail.  As with the bispectrum computational speed is vastly improved using the separable method. However, for trispectra that depend on the diagonals as well as the wavenumbers, the decomposition into separable modes is still a computationally intensive operation requiring up to $\mathcal{O}(l_{\rm{max}}^6)$ operations. Nonetheless, this decomposition need only be performed once for each model. In the particular case that the trispectra is independent of the diagonals the decomposition process may be performed efficiently in $\mathcal{O}(l_{\rm{max}}^4)$ operations. It should also be noted that the general trispectrum may be divided into contributions denoted as `reduced' trispectra. Since, for almost all theoretical trispectra presented to date in the literature, the reduced trispectra depends on five parameters (i.e. the four wavenumbers and one diagonal) a reduction in complexity for this wide range of models may also be achieved. This class of models will be discussed in a subsequent article \cite{Regan2011}.

As in the case of the bispectrum, this approach can also be used to recover trispectra from simulations and produce nonGaussian initial conditions with arbitrary trispectra for N-body codes. Once the trispectrum has been decomposed into separable modes the calculation of the trispectrum contribution to nonGaussian initial conditions is an extremely efficient operation which may be performed in $\mathcal{O}(n_{\rm{max}}^{4/3}\times l_{\rm{max}}^3)$ operations. In this paper we have also briefly outlined how the method may be extended to higher order correlators such as the quadspectra, revealing a highly efficient algorithm in the case that the quadspectrum depends only on its wavenumbers. 

The estimation of nonGaussian parameters using large-scale structure is complicated due to non-linear evolution. In this paper we have outlined some of the issues involved. The application of the separable approximation to finding the contribution to the matter density power spectrum due to the bispectrum (as well as the matter density bispectrum contribution due to the trispectrum) has been derived. In addition a prescription for parameter estimation in the fully nonlinear regime has been described.

While observational problems connected to surveys, such as because of redshift distortion and photometric errors, have not been addressed here, the generality and robustness of the methodology described in this paper suggests that a vast improvement on the scope of models investigated using large-scale structure is possible, offering a significant test of the initial conditions of the Universe. However, different large scale structure survey strategies affect the quality of the higher order correlators that can be extracted.  Given that these poly-spectra can be determined efficiently and their strong scientific motivation, this should become an issue of growing importance in survey design.

\section{Acknowledgements}

We are grateful for many informative and illuminating discussions with Michele Liguori and Hiro Funakoshi. The authors would also like to thank Emiliano Sefussati, Eiichiro Komatsu, Licia Verde, Christoph R\"ath, Holger Schlagenhaufer and Veronika Junk. EPS is grateful for the hospitality of Slava Mukhanov and Jochen Weller and the support of the DFG UNIVERSE Excellence Cluster in Munich. EPS and JRF was supported by STFC rolling grant ST/F002998/1 and the Centre for Theoretical Cosmology. DMR was supported by EPSRC, the Isaac Newton Trust and the Cambridge European Trust.

\section*{APPENDICES}

\section*{Appendix A: General Trispectrum Estimator}
In this appendix we shall elucidate in more detail the calculations involved in arriving at the expectation value of the trispectrum estimator given by equation \eqref{eq:TrispEstim}. This derivation is instructive for the calculation of many of the results presented in this paper.
\par
Similarly to the case of the bispectrum, the expectation value for the estimator is found to give
\begin{align}
\< \curl{E}\> = \frac{V}{(2\pi)^3}\int \frac{d^3\bk_1}{(2\pi)^3} \frac{d^3\bk_2}{(2\pi)^3} \frac{d^3\bk_3}{(2\pi)^3}\frac{d^3\bk_4}{(2\pi)^3} \frac{(2\pi)^6 \d_D(\bk_1+\bk_2+\bk_3+\bk_4) T^2(\bk_1,\bk_2,\bk_3,\bk_4)}{P(k_1)P(k_2)P(k_3)P(k_4)}.
\end{align}
Using the parametrisation in terms of $(k_1,k_2,k_3,k_4,K_1,K_2)$ and expanding the Dirac delta functions using \eqref{eq:deltaint} and \eqref{eq:exptobessel} we find
\begin{align}
\< \curl{E}\> =& \frac{V}{(2\pi)^3}\int \frac{(k_1 k_2 k_3 k_4 K_1 K_2)^2 d k_1 dk_2 dk_3 dk_4 dK_1 dK_2}{(2\pi)^{15}}  \frac{ T^2(k_1,k_2,k_3,k_4,K_1,K_2)}{P(k_1)P(k_2)P(k_3)P(k_4)}\nonumber\\
&\times (4\pi)^9\sum_{l_1}(2l_1 +1)\left(\int dx_1 x_1^2 j_{l_1}(k_1 x_1) j_0(k_2 x_1)j_{l_1}(K_1 x_1)\right)\left(\int dx_2 x_2^2 j_{0}(k_3 x_2) j_{l_1}(k_4 x_2)j_{l_1}(K_1 x_2)\right)\nonumber\\
&\times \left(\int dx_3 x_3^2 j_{l_1}(k_1 x_3) j_{l_1}(k_4 x_3)j_{0}(K_2 x_3)\right),
\end{align}
where the expression on the second and third lines arises from the integration over the angular variables. Next, we use the following identity from \cite{watson,9309023}
\begin{align}\label{sphbessel}
\int_0^{\infty}r^2 dr j_l(k r) j_l(k' r)j_0(\rho r)=\Theta(k,k',\rho) \frac{\pi}{4 k k' \rho}P_l\left(\frac{k^2+{k'}^2-\rho^2 }{2k k'}\right)
\end{align}
where $\Theta$ imposes the triangle condition on wavenumbers $(k,k',\rho)$ which is automatically satisfied for the trispectrum estimator at all points of the quadrilateral due to the Dirac delta functions, and $P_l$ represents the $l$th Legendre polynomial. Finally we may further simplify using the following result from \cite{Vinti1951},
\begin{align}
\sum_{l=0}^{\infty}(2l+1)P_l(x)P_l(y)P_l(z)&=\frac{2}{\pi\sqrt{g}},\qquad g=1+2xyz-x^2-y^2-z^2>0\nonumber\\
 &= 0,\qquad \mbox{otherwise}.
\end{align}
For the case of the trispectrum estimator we have
\begin{align}
x=\frac{k_1^2+K_1^2-k_2^2}{2 k_1 K_1},\qquad y=\frac{k_4^2+K_1^2-k_3^2}{2 k_4 K_1},\qquad z=\frac{k_1^2+k_4^2-K_2^2}{2 k_1 k_4},
\end{align}
and the condition $g>0$ is again satisfied for all points within the quadrilateral.

Using these expressions the expectation value of the estimator takes the following simple form
\begin{align}
\< \curl{E}\> =& \frac{V}{(2\pi)^3}\frac{1}{2\pi^4}\int_{\curl{V}_T} d k_1 dk_2 dk_3 dk_4 dK_1 dK_2 \frac{k_2 k_3 K_2}{2\sqrt{g}}   \frac{T^2(k_1,k_2,k_3,k_4,K_1,K_2)}{P(k_1)P(k_2)P(k_3)P(k_4)}.
\end{align}
In writing this expression we set $\d_D({\bf{0}})=V/(2\pi)^3$.
Therefore a suitable weight for the mode decomposition, which is a simple generalisation of the discussion in \cite{RSF10} to include an extra diagonal is given by $w(k_1,k_2,k_3,k_4,K_1,K_2)= k_2 k_3 K_2/(\sqrt{g}P(k_1)P(k_2)P(k_3)P(k_4))$. We note that the factor
$k_2 k_3 K_2/(2\sqrt{g})$ may be written as
\begin{align}\label{g1def}
 \frac{k_2 k_3 K_2}{2\sqrt{g}}=\frac{ k_1 k_2 k_3 k_4 K_1 K_2}{\sqrt{K_1^2 K_2^2 (\sum_i k_i^2 -K_1^2-K_2^2)-K_1^2 \kappa_{23}\kappa_{14}+K_2^2 \kappa_{1 2} \kappa_{3 4}-(k_1^2k_3^2-k_2^2k_4^2)(\kappa_{12}+\kappa_{34})}}\equiv\frac{ k_1 k_2 k_3 k_4 K_1 K_2}{\sqrt{g_1}},
 \end{align}
 where we denote $\kappa_{ij}=k_i^2-k_j^2$ and we denote the denominator $\sqrt{g_1}$ for brevity.

\section*{Appendix B:  Trispectrum contribution to the Bispectrum}
The contribution to the galaxy bispectrum due to the primordial trispectrum is given by
\begin{align}
B_g^T(k_1,k_2,k_3)&=\frac{1}{(2\pi)^3}\int d^3 \by  T(\bk_1,\bk_2,\by,\bk_3-\by)F_2(\by,\bk_3-\by)+\rm{2\,perms}\nonumber\\
&=\frac{1}{(2\pi)^3}\int d^3 \by d^3 \bk_4  T(\bk_1,\bk_2,\by,\bk_4)F_2(\by,\bk_4)\delta_D(\bk_4-\bk_3+\by)+\rm{2\,perms},
\end{align}
where $F_2$ is given by equation \eqref{eq:F2exp} and the permutations are cyclic in $(k_1,k_2,k_3)$. First we consider the special case that the trispectrum depends only on the wavenumbers $k_1,k_2,y,k_4$ such that we may write $T(k_1,k_2,y,k_4)=\sum_n\alpha_n q_r(k_1)q_s(k_2)q_t(y)q_u(k_4)$. The calculation is very similar to the power spectrum case and we find
\begin{align}
B_g^T(k_1,k_2,k_3)=&\sum_n \frac{\alpha_n}{4\pi^2}\frac{\sqrt{P(k_1)P(k_2)}}{(k_1 k_2)^{3/4}}\frac{q_r(k_1)q_s(k_2)}{ k_3}\int_{\mathcal{V}} dy dk_4 (y \, k_4)^{1/4} \sqrt{P(y)P(k_4)}  q_t(y)  q_u(k_4)\nonumber\\
&\times \Big[ \frac{5}{7}+\frac{2}{7}\left(\frac{k_4^2+y^2-k_3^2}{2 k_4 y}\right)^2-\left(\frac{y}{k_4}+\frac{k_4}{y}\right)\left(\frac{k_4^2+y^2-k_3^2}{2 k_4 y}\right)\Big]+\rm{2\,perms},
\end{align}
where $\mathcal{V}$ represents to domain for which the wavenumbers $(y,k_4,k_3)$ satisfy the triangle condition. The integral, we note again, may be written as a sum of products of one dimensional integrals over $y$ and $k_4$.
\par
Next we consider the more general case where the trispectrum depends also on two diagonals or equivalently the angles $\mu=\hat{\bk}_1.\hat{\bk}_2$ and $\nu=\hat{\bk}_1.\hat{\bk}_4$. In this case we may decompose the trispectrum as
\begin{align}
\frac{(k_1\, k_2\, y \,k_4)^{3/4}}{\sqrt{P(k_1)P(k_2)P(y)P(k_4)}}T(\bk_1,\bk_2,\by,\bk_4)=\sum_{n l_1 l_2}\alpha_{n l_1 l_2}q_r(k_1)q_s(k_2)q_t(y)q_u(k_4)  P_{l_1}(\mu)P_{l_2}(\nu),
\end{align}
where $n\equiv\{r,s,t,u\}$. The calculation follows much the same lines as the special case with simplification of the formulae in this case achieved using equation \eqref{sphbessel}, the following identity as described in \cite{Seaton62,BurgWhelan87}
\begin{align}
\int dx x^2 j_{l}(kx)j_{l'}(k'x) j_n(\rho x)=\Theta(k,k',\rho)\frac{\pi}{2k k' \rho^{n+1} }\sum_L Q_{nL}(k,l,k',l') P_L\left(\frac{k^2 +{k'}^2-\rho^2}{2 k k'}\right) 
\end{align}
(where the $\Theta$ function imposes the triangle condition on the three wavenumbers, $P_L$ is a Legendre polynomial and the functions $Q_{nL}$ may be found in \cite{Seaton62,BurgWhelan87}) and the identity
\begin{align}
\sum_{m_1,m_2}\left( \begin{array}{ccc}
l_1 & l_2 & L \\
m_1 & m_2 & M \end{array} \right)\left( \begin{array}{ccc}
l_1 & l_2 & L' \\
m_1 & m_2 & M' \end{array} \right)=\frac{\delta_{L L'}\delta_{M M'}}{2L+1}.
\end{align}
With these considerations we find
\begin{align}
&B^T_g(k_1,k_2,k_3)=
\sum_{n l_1 l_2}\frac{\alpha_{n l_1 l_2}}{2\pi^2}\frac{\sqrt{P(k_1)P(k_2)}}{(k_1 k_2)^{3/4}}\frac{q_r(k_1)q_s(k_2)}{k_3}P_{l_1}(\hat{\bk}_1.\hat{\bk}_2)P_{l_2}(\hat{\bk}_1.\hat{\bk}_3)\int_{\mathcal{V}} dy dk_4 (y \, k_4)^{1/4} \sqrt{P(y)P(k_4)} q_t(y)q_u(k_4)\nonumber\\
&\times \Bigg[\frac{17}{42}P_{l_2}\left(\frac{k_4^2+k_3^2-y^2}{2k_3 k_4}\right)+\frac{4\pi}{3}\sum_{l_4}\frac{(-1)^{(l_4-l_2+1)/2}h_{l_2 l_4 1}^2}{(2l_2+1)}\frac{1}{y}\left(\frac{y}{k_4}+\frac{k_4}{y}\right)\sum_L Q_{1L}(k_4,l_4,k_4,l_2)P_L\left(\frac{k_4^2+k_3^2-y^2}{2k_3 k_4}\right)\nonumber\\
&\nonumber\\
&+\frac{16\pi}{105}\sum_{l_4}(-1)^{(l_4-l_2+2)/2}\frac{h_{l_2 l_4 2}^2}{(2l_2+1)}\frac{1}{y^2}\sum_{L'} Q_{2L'}(k_4,l_4,k_4,l_2)P_{L'}\left(\frac{k_4^2+k_3^2-y^2}{2k_3 k_4}\right)\Bigg]+\rm{2\,perms}.
\end{align}

\bibliography{draft_X}

\end{document}